\documentclass[prl,aps,twocolumn,showpacs]{revtex4}

\usepackage{epsfig}
\usepackage{latexsym}
\usepackage{graphicx}

\newcommand{\bi}{\begin{itemize}}
\newcommand{\ei}{\end{itemize}}
\newcommand{\be}{\begin{equation}}
\newcommand{\ee}{\end{equation}}

\newcommand{\bea}{\begin{eqnarray}}
\newcommand{\eea}{\end{eqnarray}}
\newcommand{\beastar}{\begin{eqnarray*}}
\newcommand{\eeastar}{\end{eqnarray*}}

\begin{document}

\title{Condensation transition in DNA-PAMAM dendrimer fibers studied using optical tweezers}

\author{F. Ritort$^{\dag,\#}$, S. Mihardja$^{\ddag}$, S. B. Smith$^{\dag,\%}$ and C. Bustamante$^{\dag,\%}$\\ $\dag$ Department of Physics and Molecular \& Cell Biology, University of California, Berkeley CA 94720, USA\\$\%$ Howard Hughes Medical Institute, University of California, Berkeley CA 94720, USA\\ $\ddag$ Department of Chemistry, University of California, Berkeley CA 94720, USA\\$\#$ Department of Physics, Faculty of Physics, University of Barcelona, Diagonal 647, 08028 Barcelona, Spain\\}

\begin{abstract}
When mixed together, DNA and polyaminoamide (PAMAM) dendrimers form
fibers that condense into a compact structure.  We use optical
tweezers to pull condensed fibers and investigate the decondensation
transition by measuring force-extension curves (FECs). A
characteristic plateau force (around 10 pN) and hysteresis between the
pulling and relaxation cycles are observed for different dendrimer
sizes, indicating the existence of a first-order transition between two phases
(condensed and extended) of the fiber. The fact that we can reproduce
the same FECs in the absence of additional dendrimers in the buffer
medium indicates that dendrimers remain irreversibly bound to the DNA
backbone. Upon salt variation FECs change noticeably confirming that
electrostatic forces drive the condensation transition.  Finally, we
propose a simple model for the decondensing transition that
qualitatively reproduces the FECs and which is confirmed by AFM
images.
\end{abstract}


\maketitle

{\bf Introduction.}  In eukaryotes, genomic DNA is compacted in a
complex nucleoprotein phase known as chromatin.  Such condensation is an
important mechanism for protecting the genetic information from external
factors, as well as in storing the long DNA into a compartment with
dimensions on the order of microns.  The basic unit of chromatin
organization is a nucleosome.  There 8 histone molecules join to form a
core particle $\sim$6 nm in diameter and the DNA wraps $\sim$2 times
around that core forming a solenoid that is 10 nm wide.  When such
particles are spaced evenly along the DNA, they form the 10 nm fiber or
{\em beads on a string} structure as seen in electron micrographs.
Under various salt conditions, nucleosomes prefer higher-order condensed
structures such as a 30 nm fiber, 100 nm fiber, etc. It has long been
recognized that the main driving force for condensation is electrostatic
interaction between negative DNA and positive histone
proteins~\cite{vanHolde} but little is known about what determines the
stability of these specific structures. Do the size and shape of core
particle determine how DNA wraps around it and thus the amount of length
compaction? Are specific charge stripes on the core particles required
for DNA winding or would a uniform spherical charge distribution work as
well? What factors determine spacing between nucleosomes in the 10 nm
fiber?  In order to answer such questions it may be useful to consider
model systems that are much simpler than the DNA-histone complexes
while, at the same time, retain some of the important characteristics of
the DNA-histone fibers.

To this end, we have chosen to study the condensation of DNA by
polyaminoamide (PAMAM) dendrimers. These dendrimers are synthetic
branched polymers that exhibit such excellent properties as molecular
uniformity, well-defined size, high water solubility, and very low
toxicity.  Because of these special properties, PAMAM dendrimers are
currently being investigated for their potential in such biological
applications as the transfection of foreign genetic materials into
eukaryotic cells~\cite{kukowskabielinska}.  The PAMAM dendrimers we
investigated were synthesized via an initiator core-EDA,
ethylenediamine, followed by a series of repeated polymerization steps
(the number of steps defines the generation $G$ of the dendrimer),
terminating with amino ($NH_2$) groups.  At physiological pH values, the
surface amino groups of the dendrimers are protonated.  To study the
structure of the condensed state and to elucidate the effects of
dendrimer charge and size, we pulled DNA-dendrimer fibers using optical
tweezers.  Based on our results, we propose a simple theoretical model
for the condensation-decondensation transition that qualitatively
reproduces the force-extension curves (FECs).  This model is further
validated via atomic force microscopy (AFM) imaging of the DNA-dendrimer
complexes.  Finally, we compare our results with those of
optical-tweezers experiments done on native chicken erythrocyte
chromatin fibers\cite{CuiBus00} and on reconstituted chromatin
fibers\cite{Ben01,RowTow02}.

\begin{figure}
\begin{center}
\epsfig{file=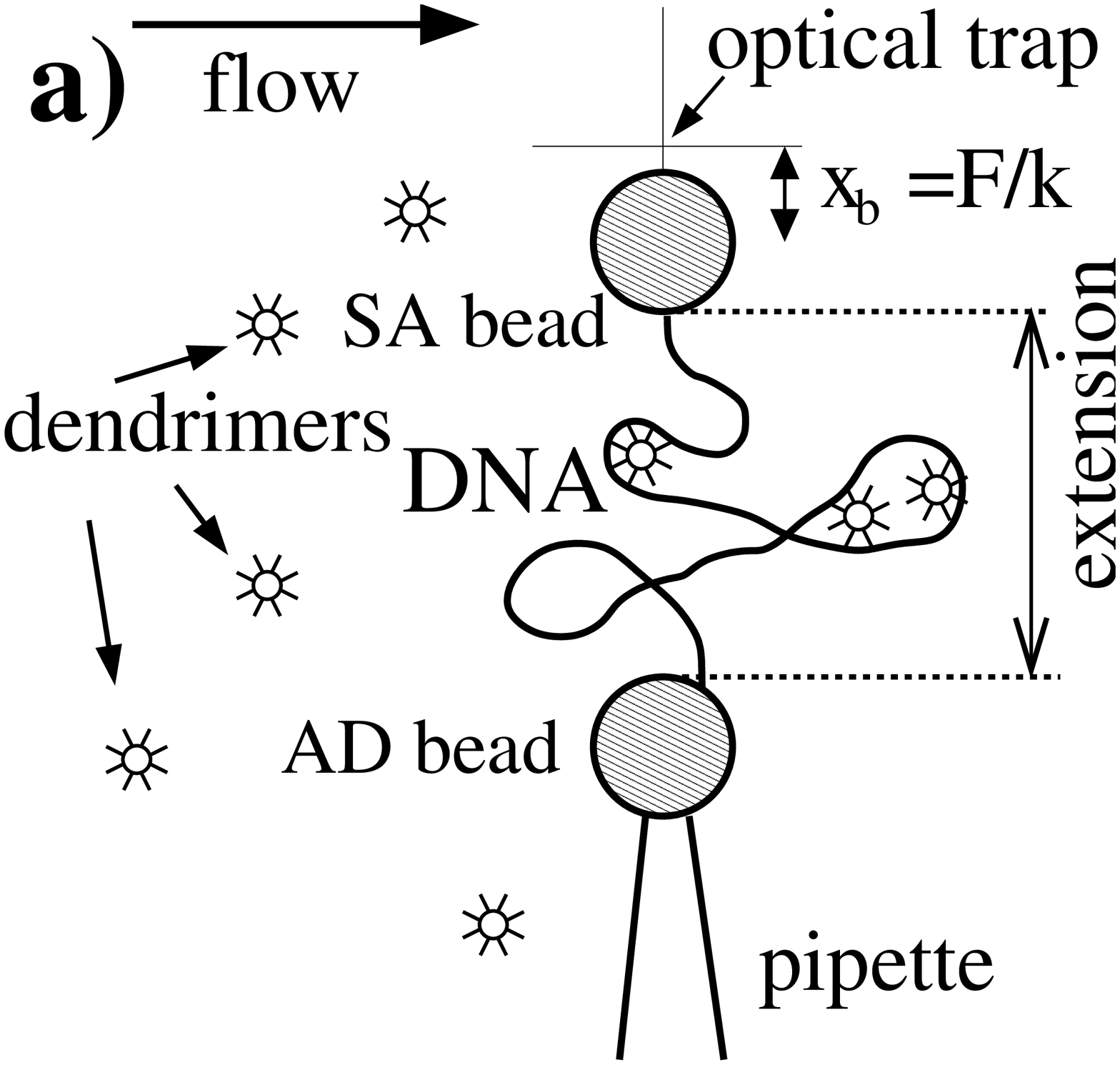,angle=0,scale=0.2}\epsfig{file=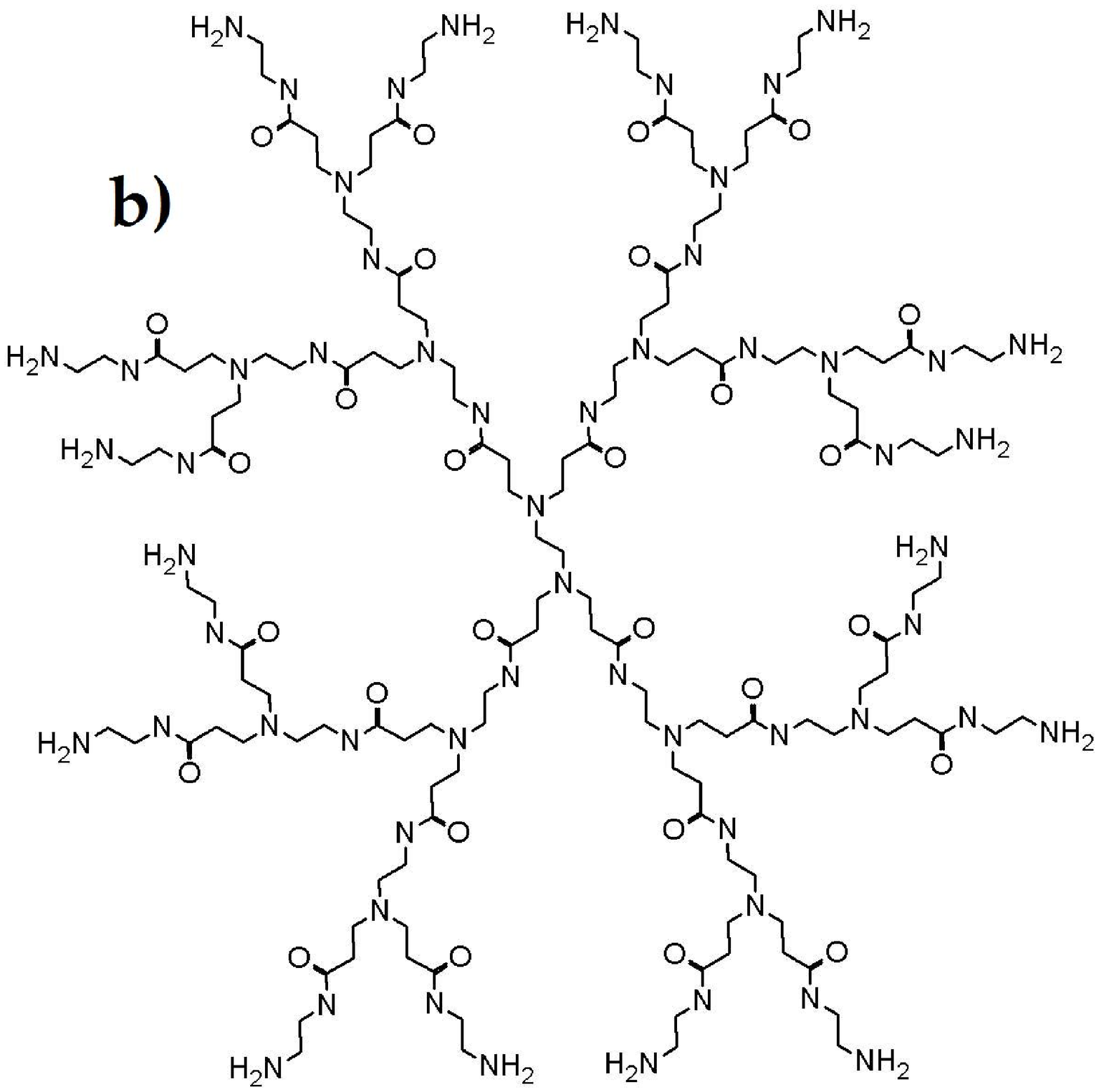,angle=0,scale=0.23}\vspace{-.0cm}
\epsfig{file=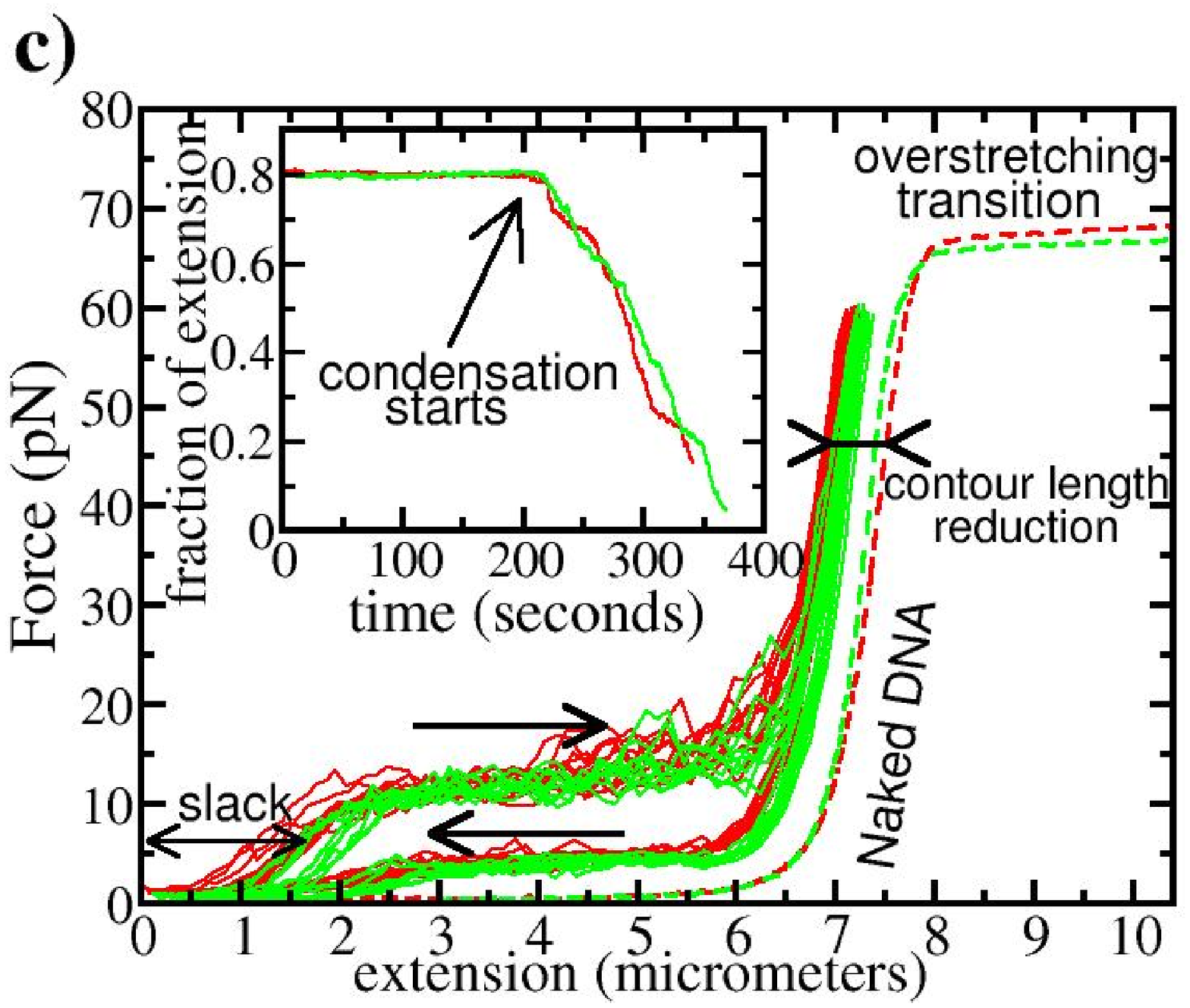,angle=0,scale=0.4}
\end{center}
\vspace{-.7cm}
\caption{(a)Experimental setup and (b) the chemical structure of a
dendrimer molecule $G2$ with 16 amino terminal groups. (c)
Reproducibility of FECs using CFP for two DNA molecules (red and green
colors) that were condensed with G8 dendrimers under the same
conditions at a constant force of 4 pN. Dashed curves correspond to
naked DNA. Inset: Relaxation curves during the condensation of the
fiber. The arrow indicates the time at which condensation starts.}
\label{fig1}
\vspace{-.4cm}
\end{figure}

{\bf Experimental setup.}  Pulling experiments were carried out with a
fragment of $\lambda$-phage DNA\cite{foot1a} and three dendrimer
generations G5, G6, G8.  Here, the number of surface amino groups
equals 128, 256, 1024 and the diameters equal 5.7, 6.7 and 9.7 nm,
respectively~\cite{foot1b}.  Most of the experiments were done in
buffer containing 100 mM of NaCl.  We used optical tweezers to exert
force on the DNA molecule as shown in Fig.~\ref{fig1}a.  Two types of
polystyrene beads (Spherotech) of 2-3$\mu m$ in diameter were used:
streptavidin (SA) coated beads and anti-digoxigenin (AD)
beads~\cite{foot1c}.  SA beads were held in a dual-beam optical
trap~\cite{SmiCuiBus02}.  A tether was made by fishing for the
free-end of the DNA with an AD bead kept on a micropipette by
suction. After first characterizing the properties of the naked DNA,
the chamber (30 $\mu l$) was rinsed with 100 $\mu l$ of buffer
containing $\sim$30 nM of dendrimers. This process typically lasted 8
min. The flow was then halted and the fiber was repeatedly pulled by
moving a piezo-controlled stage attached to the pipette.

DNA was condensed by one of two protocols: (1) constant-extension
protocol (CEP) and (2) constant-force protocol (CFP).  In the CEP, the
distance between the center of the optical trap and the bead on the tip
of the micropipette was kept constant while the chamber was rinsed
with the buffer containing dendrimers.  As the dendrimers bound to the
DNA, the force increased with time. In the CFP, a feedback loop mechanism was
used to maintain a constant force. Here, condensation was marked by a
decrease in extension of the molecule as a function of time. 

{\bf Results.}  FECs are reproducible on different molecules if they are
condensed following the same protocol and exposed to the dendrimers
under identical conditions. This effect is shown in Fig.~\ref{fig1}c
where two different DNA molecules, condensed at constant force under
identical conditions, generated the same FEC.  The FECs show the
existence of a well-defined force plateau along the pulling curve,
centered about 10 pN.  This feature is the signature of a
condensation-decondensation first-order phase transition separating a
condensed (C) from an extended (E) phase.  Upon relaxation, the fiber
once again undergoes a transition from the E to the C form that
manifests as a lower force plateau.  The FECs display the following
general features (for both protocols and for all dendrimer generations):
a) strong hysteresis between the pulling and relaxation curves; b)
repeatable pattern of FECs over successive pulling-relaxation cycles; c)
shortening of the contour length of the extended fiber ($\simeq 5\%$)
suggesting that dendrimers remain bound to the DNA in the E-phase; d)
existence of a "slack" region in the FECs at short end-to-end
extensions, where the fibers extend by $\sim1$$\mu m$ at relatively low
($<$3 pN) force. 

When the protocol conditions differ, the FECs change from experiment
to experiment. In general, for all condensation protocols, the amount
of slack is anti-correlated to the reduction observed in the effective
contour length of the fiber (i.e. the observed extension where the
force rises rapidly in the E-phase).  A larger slack is always
accompanied by a smaller reduction in contour length, suggesting that
the origin of the slack is the presence of naked DNA regions.  Using
different time protocols, we verified that the amount of slack depends
on the time exposure of the fiber to the dendrimer flow. A large slack
is observed when a fiber is exposed to flow for a short time ($\sim 5$
min.) whereas for a longer exposure ($\sim 30$ min.) the slack became
too short to measure accurately ($<$0.1 $\mu m$, data not
shown). Oddly, the measured slack decreased only while the dendrimer
buffer was flowing past the fiber but not as the buffer sat quiescent
in the chamber. After the flow had stopped, the FECs remained
repeatable in time. This effect probably indicates a rapid partition
of available dendrimers between the DNA and glass chamber walls, and a
depletion of dendrimers in the buffer. Similar depletion effects have
also been observed for positively charged proteins that stick to
negative glass walls.

To check whether dendrimers exchanged between the DNA and those in the
buffer solution during the pulling-relaxation cycle, we flowed clean
dendrimer-free buffer solution through the center of the chamber and
repeatedly pulled the fiber several times in the presence of a small
flow to {\em wash out} any dendrimers expelled from the backbone.  No
time-dependent change in the FEC was observed.  Evidently, dendrimers
are irreversibly bound to the DNA backbone, i.e. they do not bind and
unbind from the DNA during the pulling cycles. This fact was further
tested by trying to condense DNA while keeping the fiber at forces as
high as 20 pN. Condensation was observed in all cases indicating that
dendrimers bind to DNA up to very high forces. In contrast, real
chromatin fibers eventually lose their histone proteins after repeated
pulls or when subjected to high forces and eventually assume the FECs
characteristic of naked DNA~\cite{Ben01,RowTow02}.

\begin{figure}
\begin{center}
\epsfig{file=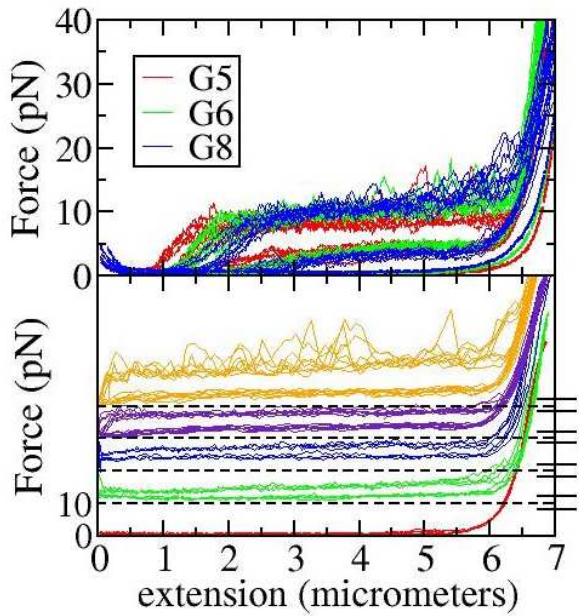,angle=0,scale=0.4}
\end{center}
\vspace{-.7cm}
\caption{Upper panel: Size dependence of FECs using CFP for different
dendrimer sizes.  The dashed curves are naked DNA.  All extensions have
been normalized to a common contour length.  Lower panel: Salt
dependence of FECs in G5 fibers. From top to bottom 10 (orange), 50 (indigo), 100 (blue), 200
(green), 500 (red) mM NaCl. For the sake of clarity FECs corresponding
to different salt concentrations have been shifted 10 pN upwards in the
figure.}
\label{fig3}
\vspace{-.4cm}
\end{figure}

For identical condensation protocols, FECs are affected by dendrimer
size in a systematic manner (Fig.~\ref{fig3} upper panel). The larger
the generation is, the higher the value of the decondensing force
plateau.  This result may be attributed to the increase in
electrostatic strength, i.e. increase in surface amino groups, with
increasing dendrimer generation.

The importance of the electrostatic interaction as the main force
stabilizing the condensed state is revealed by investigating the
effect of salt concentrations on the FECs (lower panel in
Fig.~\ref{fig3}).  Clearly, as the salt concentration increases, both
the extent of hysteresis and the value of the decondensing force
plateau decrease.  At high salt concentrations (500 mM NaCl) the
hysteresis disappears and the FEC shows a plateau at low forces ($\sim
1$ pN) similar to that observed when DNA condenses in the presence of
multivalent cations ~\cite{Baumann1}.  At low salt concentrations the
hysteresis increases noticeably and the pulling curve becomes noisy,
revealing a stickier fiber (uppermost curve in Fig.~\ref{fig3}, lower
panel).

\begin{figure}
\begin{center}
\epsfig{file=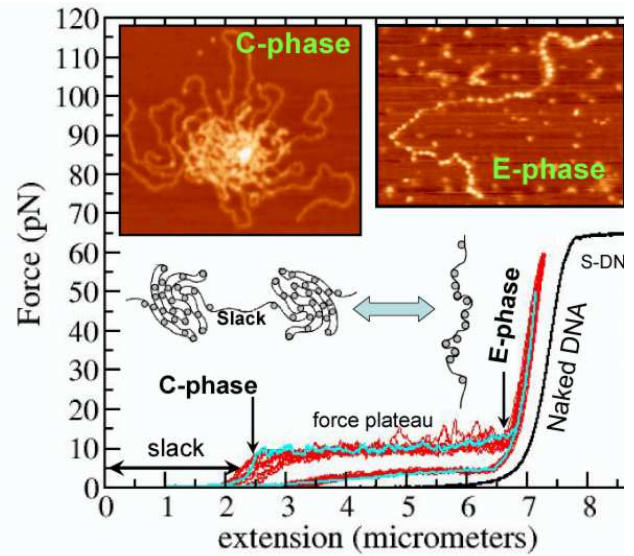,width=8.5cm}
\end{center}
\vspace{-.75cm}
\caption{Main features of FECs and schematic model for the distinct
phases (C,E) of the DNA-dendrimer fibers. The red curve corresponds to
DNA condensed with G6 in the CFP at a force of 8.4 pN. The black curve
corresponds to naked DNA. The cyan curve is the prediction of the
theoretical model (see the text). Vertical arrows denote $100\%$ of the
corresponding phases. The schematic model has been checked by taking AFM
images (G5) showing the formation of different structures (C,E-phases).}
\label{fig4}
\vspace{-.4cm}
\end{figure}

{\bf The model.}  Because DNA and dendrimers have opposite
charge, dendrimers will bind non-specifically  and will distribute
themselves non-homogeneously along the length of the DNA backbone.
In certain regions in which many dendrimers bind side by side, the
DNA tends to condense forming a compact globular structure (C-phase)
which is maintained by  electrostatic forces bridging dendrimers with
nonadjacent segments of DNA~\cite{BorNet03}.  These globular structures are
interspersed with regions of naked DNA.  When the fiber is pulled,
initially work is done to straighten the regions of naked DNA
(entropic elasticity) giving rise to the observed slack.  As these
naked regions are straightened and the force reaches a certain critical
value (the value of the plateau) the bridges that hold together the
globular structures begin to yield resulting in their mechanical
decondensation, revealed by the plateau in the FECs.  At the end of
this process, all globular structures have been converted to an
extended phase (E-phase) in which the DNA molecule resembles a line
of negative charge with a random sequence of positively charged
patches along its backbone.  Such extended structure might
conceivably resemble the 10 nm fiber in chromatin. This model and the
existence of the C,E-phases were confirmed by AFM images of the
fibers (Fig.~\ref{fig4}, insets)~\cite{foot3}.

Finally we have considered a simple theoretical description of our
model that can qualitatively reproduce the experimental results shown
in Fig.~\ref{fig1}, in particular the force plateau, the large
hysteresis observed during the pulling-relaxation cycle, and the
qualitative difference between the pulling curve (showing a plateau)
and the relaxation curve (showing a soft shoulder).  The model
consists of a collection of $N$ non-interacting two-level systems that
mimic the contacts between dendrimers and neighboring DNA segments.
The parameters describing these molecular contacts are the free-energy
of formation of each $i^{\rm th}$-contact $\Delta G_i$, its activation
barrier $B_i$ and the distance $\Delta x^{\dag}_i$ to the
contact-formed state.  Upon the action of mechanical force each
contact yields an extension $x_i$ in a thermally activated
process~\cite{Bell78}.  The resulting model shows qualitative
agreement with the experimental results by including some structural
disorder(Fig.~\ref{fig4}, blue curve)\cite{foot4}.  In general, a few
hundred contacts are needed to reproduce the experimental data
implying a similar number of dendrimers are adsorbed onto the DNA
backbone.

{\bf Conclusion.}  We have explored the behavior of DNA condensed by
PAMAM dendrimers in hopes of finding a simple model to explain chromatin
condensation. Our results do show some similarities between the
DNA-dendrimer complex and chromatin. Under certain conditions, FECs show
similar unfolding and refolding forces and have similar compaction
ratios to chromatin fibers~\cite{Ben01,RowTow02, CuiBus00}.  However, it
may be possible that nonspecific condensation of DNA by other
polycationic complexes (e.g., eukaryotic condensin,
spermidine\cite{Baumann2}) also exhibit similar behavior.

An essential aspect of chromatin structure is the fact that DNA wraps
$\sim$twice around each histone complex.  We have little evidence that
something similar happens for dendrimers.  Our AFM images cannot
resolve the path of the DNA around or through such particles. Slack in
the condensed fiber disappears with sufficient exposure to dendrimers.
Thus it probably does not represent a stable extended state like the
10 nm chromatin fiber. Previous optical-tweezers pulling experiments
on reconstituted chromatin \cite{Ben01,RowTow02} revealed a "sawtooth"
force pattern corresponding to sudden opening of the DNA wraps.
Except for a few cases (with G8), we never clearly identified such
events.  Perhaps back-folding of the terminal amino groups inside the
dendrimer and the considerable flexibility of low generation
dendrimers\cite{Maiti04} allow the dendrimers to bend around the DNA
more than the DNA bends around the dendrimers.  Such flexibility would
help explain the tight binding of dendrimers to DNA at high tension
($\leq$60 pN), where hard spheres would lose contact with a linear
molecule pulled straight~\cite{MarkoSiggia}. The structural
deformation of dendrimers bound to DNA might be confirmed by NMR, TEM
or ab-initio numerical simulations of DNA-dendrimer complexes. Further
experiments using varied salt, temperature, and pH conditions should
allow for considerable refinement of the dendrimer/DNA model. It might
be worthwhile to carry out pulling experiments with more rigid
condensing agents such as metallic gold nanoparticles to simulate
nucleosomes.

{\bf Acknowledgments.}  We acknowledge stimulating discussions with
Dr. Jean M. J. Frechet and assistance with DNA preparation by Dr. Pan
Li. This work was supported in part by the
U.S. Department of Energy grant DE-AC0376SF00098, GTL2BB "Microscopy of
Molecular Machines", the Spanish research council (FIS2004-3454) and the
Catalan government (Distinci\'o de la Generalitat).


\begin{thebibliography}{99}
\bibitem{foot1a} For the optical-tweezers experiments, a 24-kbp $\lambda$-DNA fragment(New England Biolabs) was made by digesting the linear DNA at the Xba I site. The DNA was labeled on one end with biotin and on the other end with digoxigenin.
\bibitem{foot1b} PAMAM dendrimers were purchased from Dendritic
Nanotechnologies, Inc. For the single-molecule experiments, the
dendrimers were first diluted 1/100 with water and then another 1/1000
to 10/1000 by TE buffer (10mM Tris-HCl pH 7.5, 1mM EDTA, 100mM NaCl,
1.5mM MgCl2, 0.02\%(v/v) Tween 20 and 0.01\% (w/v) milk powder).
\bibitem{foot1c} SA beads were incubated with DNA at concentrations
such that on average there was $< 1$ DNA molecule per bead. AD beads
were made by cross-linking polyclonal sheep anti-digoxigenin(Roche) to
Protein G beads (Spherotech).
\bibitem{foot3} For the AFM experiments, a 3.5 kbp DNA fragment was
incubated with dendrimers at a 4:1 (DNA phosphate:Dendrimer amine) ratio for 5
minutes at room temperature in 10mM Tris-HCl (pH 7.5), 1mM EDTA, 100mM
NaCl, 1.5mM $MgCl_2$. AFM images were obtained using the Nanoscope
IIIA Scanning Probe Microscope (Digital Instruments).
\end{thebibliography}
\end{document}